# Chaotic itinerancy and thermalization in one-dimensional self-gravitating systems

Toshio Tsuchiya*
*Department of Astronomy, Kyoto University, Kyoto, 606-01, Japan*

Naoteru Gouda
*Department of Earth and Space Science, Osaka University, Toyonaka, 560, Japan*

Tetsuro Konishi
*Department of Physics, Nagoya University, Nagoya, 464-01, Japan*

## Abstract

This is the third paper of the series of our studies of the one-dimensional self-gravitating many-body systems. In our first paper (Paper I), we found there are two stages in the relaxation of the water-bag to isothermal distribution, the microscopic relaxation and the macroscopic relaxation as the faster and the slower relaxation, respectively. In the second paper (Paper II), we derived the time scales of the relaxations numerically, and discussed the mechanisms, for the water-bag distribution. After these paper, we found phenomena that the system returns from the isothermal distribution to the water-bag. Thus the time scale of the macroscopic relaxation is found not to be sufficient for true thermalization.

In this paper, we thus study the transition phenomena after the first transition from a quasiequilibrium. We found that irrespective of the initial conditions, the system wanders between many states which yield the same properties of the quasiequilibrium as the water-bag. This itinerancy is most prominent in the time scale $t \sim 10^5 \sim 10^6\, t_c$. In the midway between two succeeding quasiequilibria, the system experiences a transient state, where one particle keeps exclusive high energy and its motion decouples with the others. Though the transient state is not the thermally relaxed equilibrium, its distribution quite resembles the isothermal distribution. Thus the macroscopic relaxation we discussed in the previous papers corresponds to the transition from one of quasiequilibria to a transient state. Averaging the behavior over a time scale much longer than the macroscopic relaxation time gives the isothermal distribution. Distribution of lifetime $\tau$ of the transient states yields a power-law distribution of $\tau^{-2}$. The transient state gives a clear example of chaotic itinerancy in conserved dynamical systems. The mechanism of the onset of itinerancy is examined.

*Electronic address: tsuchiya@kusastro.kyoto-u.ac.jp

## I. INTRODUCTION

In this paper, we study the evolution of the one-dimensional self-gravitating many-body system. Originally this system was introduced as a model of the motion of stars in a direction normal to the disk of a highly flattened galaxy [1,2]. In 1970s and 1980s this system was used to study mechanism of relaxation and evolution of galaxies. The problem was that the age of galaxies are far short to the relaxation time driven by stellar encounters [3]. Lynden-Bell suggested a mechanism of relaxation effective in a time scale of several crossing time (typical time scale for a star to traverse the galaxy). This mechanism is called the *violent relaxation*. Many works were devoted to investigate the mechanism (see, e.g., [4–6] and references therein), and it was revealed that the violent relaxation does not work completely for the thermalization.

Severne and Luwel [7] suggested that there are three phases in relaxation. If the initial state is far from equilibrium, violent oscillation of the mean field gives rise to the violent relaxation [8] for the first several oscillations. This phase is called the *virialization phase*. After the system is almost virialized, remaining small fluctuations of the gravitational field cause the change of the individual particle energies. They called this era the *collisionless mixing phase*. After that, the *collisional relaxation phase* takes place, in which the particle interactions tend to drive the system towards the microscopic thermal equilibrium. This is the thermal evolution.

Succeeding numerical investigations showed, however, some complex phenomena. Some initial conditions yielded very long lived correlations, which indicate the incomplete relaxation [9]. Some authors tackled this issue in the view of chaotic dynamics [10–13]. Many other systems show a tendency that increasing the number of degrees of freedom leads the systems more chaotic. The gravitating systems, however, have a different character; Reidl and Miller showed that the system becomes most chaotic for $N \sim 30$ [14], and larger systems are less chaotic. This different characteristics is a possible cause of incompleteness of relaxation in the long time scale. Therefore it is necessary to study chaotic aspects in the gravitating systems.

Recently we noticed that some phenomena found in the field of chaotic dynamics have close connection with the dynamics of gravitating systems. For example, Konishi and Kaneko [15] studied structure formation in a symplectic coupled map system, and found that one of quasiequilibria, the clustered state, shows the ordered motion which comes from sticking chaotic motion around self-similar structure of remnants of resonance islands in the phase space. Their work gave an example that coexistence of different chaotic regions in phase space is essential to dynamical structure changes and long range force can create an ordered state from chaotic dynamics. Another example is the famous FPU problem [16], in which energy distribution does not approach the equipartition for a long time. In this case, this *induction phenomena* [17,18] is considered as motions near KAM tori, which imitate the ordered motion. These two examples suggest that variety of relaxation process found in 1-dimensional self-gravitating systems can be considered as a dynamical problem which is related to complex structures in phase space.

Usual three-dimensional $N$-body system is hard to study such a long course of evolution because of some technical difficulties such as the effect of softening and a long computational time.

One-dimensional systems are superior in this problem, because the gravitational force in one-dimension is uniform, thus the equations of motion are reduced to algebraic equations rather than differential equations. That allows us to compute the evolution numerically in a very high accuracy and for a long time.

This is the third paper of a series of our studies on the relaxation process of one-dimensional self-gravitating systems. Combinational approach of the macroscopic and microscopic dynamics have revealed the long time relaxation process to the thermal equilibrium. Here we briefly summarize our previous works on this subject.

In Paper I [19], we showed that for the case of the water-bag initial conditions, which has a uniform density in phase space, there exist two characteristic time scales even after the virialization phase. The shorter one is the *microscopic relaxation*, which means that mixing among energies of particles has developed completely as the result of mutual interactions but the global energy distribution is not transformed into that of the thermal equilibrium, which is the isothermal distribution. This transformation of the global distribution occurs in a much longer time scale, which we refer to as the *macroscopic relaxation time*.

In Paper II. [20] we investigated the precise time scale and the physical mechanism of the two relaxations. From the numerical analyses we revealed the course of evolution of the water-bag distribution: A phase point which describes the state of the system moves like a random walk in phase space. This random motion corresponds to the diffusion in the phase space, and the time scale of the diffusion gives the time scale of the microscopic relaxation, which is about $N$ times the crossing time. This diffusion is the mechanism of relaxation in usual molecular dynamics. If there is no other physical process, the water-bag distribution transforms into another, possibly the isothermal distribution in the time scale of this diffusion time scale. However, in the case of the self-gravitating systems, there seems to be some barriers which restrict the phase point in the water-bag region in phase space for a long time. The transition from the water-bag to the isothermal distribution happens stochastically when the phase point find a small window on the barrier to escape from. The macroscopic relaxation has the much longer time scale $4 \times 10^4 N$ times the crossing time.

In actual fact the macroscopic relaxation time scale was defined as the length of time when the first transition of the macroscopic distribution started. As explained in Sect. II, we found that there was the second transition at which the system returned to the quasiequilibrium and after that transformed into the isothermal distribution and this repeated again and again. Therefore we need to make the process of true thermalization clear. Besides, the mechanism of the transition of the macroscopic distribution is investigated in this paper.

Section II gives a description of our model and the method of analyses. In Sect. III the results of numerical simulations are shown and in Sect. IV the interpretation of the results is presented. Conclusions and discussions are given in Sec. V.

## II. MODEL

The system comprises $N$ identical mass sheets, each of uniform mass density and infinite in extent in the $(y, z)$ plane. We call the sheets *particles* in this paper. The particles are free to move along the $x$ axis and accelerate as a result of their mutual gravitational attraction. The Hamiltonian of this system has the form

$$H = \frac{m}{2}\sum_{i=1}^{N} v_i^2 + (2\pi G m^2)\sum_{i<j}|x_j - x_i|, \tag{1}$$

where $m$, $v_i$, and $x_i$ are the mass (surface density), velocity, and position of $i$th particle respectively.

Since gravity has no scale of length, our numerical results can be scaled by the total mass $M = Nm$, and the total energy $E$. The typical length $L$, velocity $V$, and the crossing time $t_c$, which is typical time for a particle to traverse the system are expressed as follows:

$$L = (4E)/(4\pi G M^2), \tag{2}$$

$$V = (4E/M)^{1/2}, \tag{3}$$

$$t_c = (1/4\pi G M)(4E/M)^{1/2}. \tag{4}$$

Following the convention, we choose these values as units to measure the system, by setting $M = E = 4\pi G = 1$.

Rybicki [21] derived the distribution of the thermal equilibrium from the condition that the system is ergodic and has the maximum entropy. In the limit that $N \to \infty$ the distribution function is

$$f(\varepsilon) = \frac{1}{8}\left(\frac{1}{2\pi}\right)^{1/2}\left(\frac{3M}{2E}\right)^{3/2}\exp\left[-\frac{3M}{2E}\varepsilon\right], \tag{5}$$

where $\varepsilon \equiv \frac{v^2}{2} + \Phi(x)$ is the specific energy (energy per unit mass), and

$$\exp\left[-\frac{3M}{2E}\Phi(x)\right] \equiv \mathrm{sech}^2\left(\frac{x}{8E/3}\right). \tag{6}$$

This distribution is often called the *isothermal distribution.* For our case, the number of particle ($N = 64$) is large enough for the system to be described by the isothermal distribution.

Individual particles change their energy along the temporal evolution [19]. These individuals are distinguished by the indices $i = 1, \ldots, N$. We call them the *named particles*, and their energies are denoted by $\varepsilon_i(t)$. Now we introduce another set of energies, $\varepsilon^{(\mu)}(t)$, $\mu = 1, \ldots, N$, which are the $\mu$-th lowest energy at that instant, then $\epsilon^{(\mu)} < \epsilon^{(\nu)}$ for $\mu < \nu$. $\mu$ is referred to as *sorted indices*, and $\varepsilon^{(\mu)}$ sorted energy.

The variation of $\varepsilon^{(\mu)}$ shows the evolution of the global distribution in addition to the microscopic dynamics. We have employed the cumulative energy distribution $\hat{\nu}(\varepsilon)$ to visualize the global distribution:

$$\hat{\nu}(\varepsilon) \equiv \frac{1}{N}\left\{\text{the number of particles with the energy smaller than } \varepsilon\right\}. \tag{7}$$

Practically it is expressed by the sorted indices

$$\hat{\nu}(\varepsilon^{(\mu)}) = \frac{\mu}{N} \tag{8}$$

From the analytic solution of the isothermal distribution (eq. 5), the expectation value of $\varepsilon^{(\mu)}$, which we referred to as $\bar{\varepsilon}^{(\mu)}$, is given by

$$\hat{\nu}(\bar{\varepsilon}^{(\mu)}) = \int_{\varepsilon_{min}}^{\bar{\varepsilon}^{(\mu)}} f(\varepsilon) dx dv = \frac{\mu - 1/2}{N}. \tag{9}$$

The real system exhibits thermal fluctuation. Thus $\varepsilon^{(\mu)}$ fluctuates around its expectation value $\bar{\varepsilon}^{(\mu)}$ in time. The amplitude of the fluctuation, $\Delta\varepsilon_{\pm}^{(\mu)}$, can be estimated as follows:

$$\frac{1}{2N} = \hat{\nu}(\bar{\varepsilon}^{(\mu)}) - \hat{\nu}(\bar{\varepsilon}^{(\mu)} - \Delta\varepsilon_{-}^{(\mu)}) = \hat{\nu}(\bar{\varepsilon}^{(\mu)} + \Delta\varepsilon_{+}^{(\mu)}) - \hat{\nu}(\bar{\varepsilon}^{(\mu)}) \tag{10}$$

In this way, each range between $\Delta\varepsilon_{-}^{(\mu)}$ and $\Delta\varepsilon_{+}^{(\mu)}$ holds mass of a particle. In the case of the highest energy particle, $\Delta\varepsilon_{+}^{(\mu)}$ becomes infinity. For the sake of convenience, we choose $\Delta\varepsilon_{+}^{(\mu)}$ as the upper boundary of the energy in side which 95% of mass is contained.

Figure 1 shows the analytic values $\bar{\varepsilon}^{(\mu)}$ by filled diamonds for $N = 64$. The error bars indicate the variances $\Delta\varepsilon_{\pm}^{(\mu)}$. Later we use these values to identify the distribution as the isothermal.

## III. ITINERANT BEHAVIOR AMONG QUASIEQUILIBRIA, AND TRANSIENT STATES

In the Paper I and II, we studied the relaxation of the water-bag initial distribution. Figure. 8 and 9 in Paper I showed the evolution of the global distributions in a long time scale. In order to make statistical analysis of the macroscopic relaxation time, we carried out a large number of simulations of the statistical ensemble of the water-bag initial distribution in Paper II. Then we derive the time scale of the macroscopic relaxation by analyzing the first transition time from the water-bag to the isothermal distribution. At that time we did not noticed longer time evolution after the transition, but later we found that in some runs the system returned from the isothermal distribution to the water-bag. Figure 2 shows an example of the recurrent transition between the water-bag and the isothermal-like distribution. At $t \sim 4.8 \times 10^6$ the distribution still remains in the water-bag distribution (denoted by the open diamond, $\diamond$), and at $t \sim 5.8 \times 10^6$ it transforms into the isothermal-like distribution ($\bullet$). After that the system returns to the water-bag at $t \sim 6.9 \times 10^6$ ($\square$). Thus this isothermal-like distribution is not the true thermal equilibrium, but some temporal state.

In Paper II, we asserted that transition of the global distribution from the water-bag to the isothermal distribution can be considered as escape of the phase point in the phase space from a barriered region. In order to understand true thermalization, longer time evolution after the transition need to be investigated. We take 64 initial conditions randomly taken from the isothermal distribution. The number of particles is restricted into 64, because we have difficulty in calculating such a long time evolution of larger system. However, the number, 64, is larger than the critical population found by Reidl and Miller [14], and the evolution of larger systems can be extrapolated from our results.

The initial conditions are taken such that the particles are randomly distributed with the probability distribution of the isothermal. Since the isothermal distribution of the position and velocity is defined by projection of the microcanonical ensemble, this choice of the initial conditions give uniform sampling of the points in the phase space. We have examined many such initial conditions and found that the manner of evolution of various runs is almost the same in the time scale longer than $10^8$. Figure 3 shows one of the results as an example. The curves indicate averaged time variation of the sorted energies, which are derived by the following procedure:

$$\langle \varepsilon^{(\mu)} \rangle (t) \equiv \frac{1}{\Delta t} \int_{t-\Delta t}^{t} \varepsilon^{(\mu)}(t) dt, \tag{11}$$

where $\mu = 1$ is the lowest energy and $\mu = 64$ the highest. In this figure $\Delta t \sim 2 \times 10^6$, and the maximum time scale is $5 \times 10^8$. This time scale is much longer than the cases we examined in the previous papers and in Fig. 2. In this time scale variations of the averaged energies are not quite large and the values suit the analytic values in the isothermal distribution. Therefore this state is considered to be the thermal equilibrium and ergodic. This fact also clear in the variation of $\Delta(t)$ in Fig. 4, where

$$\Delta(t) \equiv \varepsilon_0^{-1} \sqrt{\frac{1}{N} \sum_{i=1}^{N} \left[ \overline{\varepsilon_i}(t) - \varepsilon_0 \right]^2}, \tag{12}$$

and

$$\varepsilon_0 \equiv \lim_{T \to \infty} \frac{1}{T} \int_0^T \varepsilon_i(t) dt = 5E/3, \tag{13}$$

and $\overline{\varepsilon_i}(t)$ is the averaged value of the energy of $i$-th named particle until $t$. As discussed in Paper I, the time dependence that $\Delta(t) \propto t^{-1/2}$ denotes that the fluctuation of $\varepsilon_i(t)$ is similar to the noise in the thermal equilibrium. We found that the system is in the thermal equilibrium in the time scale longer than $10^7$ in Fig. 4. In the shorter time scale between $10^5$ and $10^7$, the violent variation of $\Delta(t)$ suggests some non-stationary phenomena turn out.

The variations of $\langle \varepsilon \rangle^{(\mu)}(t)$ in the time scale $10^6$ are shown in Fig. 5. The variations are locally averaged over $\Delta t \sim 2 \times 10^3$. In this time scale the variations seem not merely the thermal fluctuations but indicating existence of two distinctive states. One is the state found in the regions, e.g., $0 < t \lesssim 10^5$ or $1.4 \times 10^5 \lesssim t \lesssim 5.5 \times 10^5$, which are separated by the other states, seen as prominent peaks. The first state shows stable variation of $\langle \varepsilon \rangle^{(\mu)}$ like a quasiequilibrium. To make it clear, fine-grained evolution of the first $10^5$ of Fig. 5 is shown in Fig. 6. The averaging interval $\Delta t \sim 200$. There is no prominent peak in this figure. The fluctuations of energies originate in the thermal noise, which is evident in Fig. 4, where $\Delta(t)$ decreases as $t^{-1/2}$ until $t \sim 10^5$. Figure 7 shows another evidence that mixing of energies occurs well. The thick solid line indicates the variation of the energy of the highest energetic (sorted) particle, $\mu = 64$. Analytic values of $\bar{\varepsilon}^{(\mu)}$ and $\Delta \varepsilon_{\pm}^{(\mu)}$ for the highest energetic particle are also shown by dash-dotted lines. The two named particles each becomes the maximum, sometimes, but after a short while, say several 100, it loses its energy, and then another particle takes place. These facts means this state is in the thermal equilibrium in the microscopic point of view. However, the energy distribution averaged over $10^5\,t_c$ is slightly but certainly different from the isothermal distribution (Fig. 8). Therefore this state is one of the quasiequilibria we discussed in the previous papers. In the previous papers, we treated only the water-bag distribution, but the present results are evidence of existence of many other quasi-equilibria.

The other state looks like prominent peaks of the highest energy found at $t \sim 1.2$, 5.9, 6.8, $9 \times 10^5$. In this state, difference of energies between the highest and the second highest is much larger than in the quasiequilibria. In fact, in the this state, only one particle bears the highest energy throughout its lifetime. Figure 9 shows the evidence, where the thick solid line indicates the variation of the highest sorted energy, and the thin lines are two of the named particles. The named particle 1 is responsible for the highest peak at $t \sim 9 \times 10^5$, and the named particle 2 for the second highest peak at $t \sim 6.8 \times 10^5$. The other two peaks are also contributed by single named particles. We refer to this state as the *transient state.*

### A. Analyses of the transient states and the quasiequilibria

Here we show results of several analyses in order to identify the transient state.

Since the difference between the quasiequilibrium and the transient state is most prominent in the behavior of the highest energetic particle, we first examined its properties. As the particles vary their energy in time, they become the highest energetic particles in turn. The intervals between two succeeding exchange of particles, $\tau$, is referred to as the "lifetime" of the highest energetic particle to brevity. Figure 10 shows the statistical relation between the lifetime and the averaged energy of the highest energetic particles. The energy indicated in the vertical axis is averaged over its lifetime. These data are taken from one of the simulations calculated until $t = 10^7$, but the results did not depend on the initial conditions. Each dot shown in the figure corresponds to a different part of time in the evolution which includes the periods of both the quasiequilibria and the transient states. There are two noticeable feature in this figure. One is the concentration of the particles on the line at $\tau \sim 100$. The other is the branch which extends from $\tau \sim 100$ to $\tau \sim 10^5$. The time scale of the former branch ($\tau \sim 100$) corresponds to the microscopic relaxation time ($\sim N\,t_c$, see Paper II). This branch originates in states in which energies are well mixed, and thus it belongs to the quasiequilibria. On the other hand, the latter branch distributed from the root of the quasiequilibrium branch, extending straightly to $\tau \sim 10^5$ and $\langle \varepsilon \rangle \sim 1.4$. The particle with higher energy tends to last longer. However, in the quasiequilibrium branch, there exist the particles with comparably high energy. The difference of the lifetimes depends on the difference of the energies between the highest and the second highest particle, which is proved in Fig. 11. These facts tell us that appearance of exclusively high energetic particle is the origin of the transient state. Now let us understand the mechanism to accelerate and decelerate it, which governs the transition of states.

Figure 12 and 13 are the distributions of position and velocity in one-body phase space ($\mu$-space) of a quasiequilibrium and the beginning of a transient state, respectively. In the quasiequilibrium the distributions are quite close to virial equilibrium, which means that the distribution is "round" in the $\mu$-space and then it is stationary as the particles rotate. On the other hand, at the beginning of the transient state, the distribution shows elongation (in Fig. 13 to the direction of the velocity axis). It means that the system oscillates in the real space. Three particles are marked by different symbols, and the particles $i = 9$ is the highest energetic particle in the succeeding transient state. Particles rotate clockwise around the origin, thus the particle, $i = 9$, is the one which is located at the most outside and has delayed phase against the oscillation of the rest of the system. The delay of phase causes the resonant acceleration of the particle. (The mechanism of the acceleration was studied in ref [6].) The oscillation of the global distribution is considered as a kind of thermal fluctuation. It is well known that self-gravitating systems have many proper modes of global oscillation [22], which means the oscillation arising even in the limit of $N \to \infty$. By the thermal fluctuation, sometimes some mode of oscillation is excited, and at that very instance a particle located at the most outside and in delayed phase to the oscillation, e.g., the position of the particle, $i = 9$, in Fig. 13, is suffered resonant acceleration. The process of losing energy at the end of transient states is the inverse of that of gaining energy.

Figure 14 is fine-grained view of evolution of one of the transient states. The maximum time scale is $10^5$ and $\Delta t \sim 200$. The sorted particle $\mu = 64$ has exclusively high energy and it is composed of only one named particle until $t \sim 9.2 \times 10^4$. As for the rest particles, $1 \leq \mu \leq 63$, energies are well mixed as if they form a quasiequilibrium of $N = 63$. In the transient states the energy of the highest energy particle ($\mu = 64$) becomes an approximate conservative quantity in the system.

As clearly seen in Fig. 5, the system wanders between the quasiequilibria and the transient states. Although a transient state is not the true thermal equilibrium, its distribution quite resemble the isothermal distribution. Figure 15 shows the energy distribution of one of the transient states, which is averaged over its lifetime. The macroscopic relaxation time we determined in Paper II is the first transition time from a quasiequilibrium, in other words the life time of the quasiequilibrium. The true thermal equilibrium is accomplished by averaging the variations over a time scale longer enough for the system to experience all quasiequilibria, and all the transient states. From our numerical experiments it is considered an order longer than that we determined in Paper II, thus the correct time scale $T_{\text{thermal}} \sim 4 \times N t_c$.

## IV. INTERPRETATION OF THE RESULTS

In paper II, we showed that there seems to be a barriered region in phase space and the phase points inside the region realize the water-bag distribution. A phase point which starts in the region travels all over the inside and thus the water-bag distribution becomes a quasiequilibrium. We show in the present paper that the water-bag distribution is not the special case but there are many such quasiequilibria besides the water-bag.

One of the new results in this paper is the discovery of the transient state, which have an energy distribution resemble to the isothermal distribution, but do not relax even in microscopically. In that states the particle which has maximum energy moves differently from the others, and its energy is roughly constant in time. In this case, the energy of the most energetic particle (hence the sum of energies which the rest of the particles have) can be considered as an additional conservative quantity and thus the orbit in the phase space is restricted in the hyper surface which is determined by the conservative quantity (Fig. 16).

Furthermore we found that the system wanders between the quasiequilibria and the transient states, and the averaging over the time scale of the itinerancy gives the isothermal distribution, in which the distribution of particles is well mixed.

From these facts we will picture how the evolution goes on. Figure 17 shows the basic idea of evolution in phase space. There are many barriered region in the phase space, which are denoted by dashed circles in Fig. 17. The barriers which enclose the regions obstruct the orbit to go out of the region, then the orbit travels all over the region ergodically. The barriers are, however, not complete, and the orbits escape through small windows on the barriers. The escaped orbits enter the transition stage. In this stage the number of dimension of the orbits reduces as if the orbits are passing through a narrow corridor. And the orbits enter another barriered region and the system becomes another quasiequilibrium, again. The orbits repeat the same itinerancy again and again, and at last, the orbits travels all over the phase space. In this stage, the thermal equilibrium is established.

In order to obtain more information about the structure of the orbits in the phase space, we analyzed the probability distribution of the lifetime of the transient states and the quasiequilibria. As discussed in the previous section, in the quasiequilibria, the highest energetic particle loses its energy at most in several hundred crossing times. Hence we empirically define the transient state as a period of a highest energetic particle with a lifetime longer than $1000\,t_c$, and a quasiequilibrium as a period between two transient states. Figure 18 and 19 are the probability distribution of the lifetime $P(\tau)$ of the transient states and the quasiequilibria, respectively. The lifetime of transient state has clear single power law distribution, $P(\tau) \propto \tau^{-2}$. This power law dependence extends to the limit of detection in our restricted time of integration. On the other hand, for the quasiequilibria, the probability distribution has different feature. It has the power law component as well in the region $\tau \lesssim 10^5$, but also sharp cutoff at $\tau \sim 10^6$. The basic features of the lifetime distributions do not change by modifying the threshold value which discriminates transient state and quasiequilibrium. Recalling the result in Paper II that the lifetime of the water-bag distribution has an exponential distribution, it is possible that other quasiequilibria also have a typical time scale. Nevertheless, there should be variety of their lifetime, thus Fig. 19 should be interpreted as the distribution of the lifetime among quasiequilibria. The power law distribution is possibly an aspect of fractal structure of the orbits in the phase space. The cut off in the distribution shows existence of upper limit of the lifetime. The longest lived quasiequilibrium is something like the water-bag distribution, which has a typical life time, $3 \times 10^7\,t_c$, for $N = 64$.

## V. CONCLUSIONS AND DISCUSSIONS

As conclusion we summarize the evolution of the one-dimensional self-gravitating many-body systems, based on the results which are obtained by the series of our works.

We suggest that there are roughly four stages of evolution in the system.

1. Virialization phase: The system which is far from the virial equilibrium experiences violent oscillation of the mean field. The violent relaxation leads the system to one of the virial (dynamical) equilibrium at the end of this stage. Many works were devoted to the study of this stage in 1970s and 1980s. In general, this violent oscillation ceases in several $t_c$.

2. Dynamical equilibria: This stage corresponds to the era with the time scale of several $t_c \lesssim t \lesssim N t_c$. The system is in a dynamical equilibrium such as the water-bag distribution, which we investigated in Paper I and II. In this stage the energies of the individual particles conserve. Small fluctuations of mean field are driving the system to the equipartition, but it is not yet effective in this time scale.

3. Quasiequilibria: After the change of energies driven by the fluctuation of the mean field becomes effective, the energy distribution achieves the equipartition. This process of relaxation to equipartition is called the microscopic relaxation. The time scale of the microscopic relaxation is about $N t_c$ irrespective of the initial conditions and after that the system is settled in a quasiequilibrium. This quasiequilibrium is realized by the orbit which is restricted in a part of the phase space. This region is enclosed by some kind of barrier which obstructs the orbit to go out of the region, thus the orbit eventually travels all over the barriered region and exhibits the nature of equilibria. That is why the mixing of energies occurs sufficiently but the global distribution does not change. The water-bag distribution is one of the quasiequilibria, and we found in this paper that there exist many similar quasiequilibria. The orbit which is initially located in one of the regions stays inside. It escapes from the region when it find a window to the outside. This happens stochastically, but the typical time scale is, in the case of the water-bag, $4 \times 10^4 N t_c$. As for other quasiequilibria, the time scale is about the same. When the orbit escapes from one of the barriered region, the global distribution transforms. This process is, we call, the macroscopic relaxation.

   The microscopic relaxation and the collisionless mixing, which Severne and Luwel discussed, are considered to be different aspects of the same processes. The collisionless mixing is defined in the system with two species, as the era that energies of both light and heavy particles change and mix well, but equipartition of energies between light and heavy particles is not attained. Since we use one component system in our papers, we cannot define this kind of collisionless mixing, but equipartition among the same species. In the case of two components, however, the equipartition between light and heavy particles leads a structual change of the system. Therefore in the sense that the mixing of energies occurs without macroscopic transformation, the collisionless mixing corresponds, or at least is a initial part of the microscopic relaxation.

4. The thermal equilibrium: In this paper, we found that the system in a quasiequilibrium transforms into another quasiequilibrium, experiencing a transient state in the midway between the quasiequilibria. In the transient state one particle holds exclusively high energy, and its motion decouples from the others. Since the energy of the highest particles is approximately a conservative value, the orbit in the phase space is restricted in a hyper surface (see Fig. 16(b)). The system wanders between the quasiequilibria and the transient states. Averaging this behavior over a time scale much greater than the macroscopic relaxation time, gives the isothermal distribution. In this time scale, the system becomes ergodic and then the thermal equilibrium.

In this paper we examined the relaxation process of classical 1-dimensional self-gravitating system, and found that relaxation from water-bag to isothermal distribution, which is believed to be the thermal equilibrium, is not straightforward. On the contrary the system has many quasiequilibria and experiences complex itinerant procedure from one quasiequilibrium to another. The system can stay in one of the quasiequilibrium for long time, hence one can observe the system in one of the quasiequilibria, not in the true equilibrium.

The system wanders among quasiequilibria not by external force or noise, but by its own dynamics. This kind of dynamic wandering among macroscopically different states is known as *chaotic itinerancy* [23], whose examples are found in brain, nonlinear optics, water molecules, and so on. Our system is a clear example of chaotic itinerancy. Moreover our system is unique in that we have clarified the transition process, as is described in section III A.

Since our system, 1-dimensional system with long range force, is quite simple, our analysis would be useful to understand chaotic itinerancy in conserved systems.

At the view of astrophysics, the time scales in which the state of the system changes are most concerned. For example, elliptical galaxies have the typical crossing time of $10^8$ years, and the thermalization time scale is supposed to be $10^{17}$ years. The age of elliptical galaxies is order of $10^{10}$ years, thus the present state is in the middle of them. Although there are big differences between our one-dimensional system and the real one, our results gives a stimulating implication that the elliptical galaxies are evolving under another relaxation mechanism with intermediate time scale. Kandrup [24] discussed the possible mechanism of evolution which is related with the structural instability in the boundary between regular and stochastic orbits, in much shorter than thermalization time. Therefore examination of the evolution with intermediate time scale between the crossing time and the thermalization by three-dimensional $N$-body simulations should be the next subject.

## ACKNOWLEDGMENTS

The authors thank: Y. Aizawa for giving us stimulating comments and motivating us to study this subject; S. Inagaki and K. Kaneko for encouraging us; B. N. Miller and C. J. Reidl for useful discussions. TT was supported by JSPS Research Fellow. The computation was carried out on Hewlett-Packard HP730 of the theoretical astrophysics division, National Astronomical Observatory, Japan.

## REFERENCES

[1] J. H. Oort, Bull. Astr. Inst. Netherlands **6**, 289 (1932).
[2] G. Camm, Mon. Not. Roy. Astron. Soc **110**, 305 (1950).
[3] S. Chandrasekhar, *Principles of Stellar Dynamics* (University of Chicago Press, Chicago, 1942).
[4] T. de Zeeuw and M. Franx, Annu. Rev. Astron. Astrophys. **29**, 239 (1991).
[5] Y. Funato, J. Makino, and T. Ebisuzaki, Publ. Astron. Soc. Jpn **44**, 291 (1992).
[6] T. Yamashiro, N. Gouda, and M. Sakagami, Prog. Theor. Phys. **88**, 269 (1992).
[7] G. Severne and M. Luwel, Astrophys. Space Sci. **122**, 299 (1986).
[8] D. Lynden-Bell, Mon. Not. Roy. Astron. Soc **136**, 101 (1967).
[9] B. N. Miller and C. J. Reidl, Jr., Astrophys. J. **348**, 203 (1990).
[10] C. Froeschlé and J.-P. Scheidecker, Phys. Rev. A **12**, 2137 (1975).
[11] G. Benettin, C. Froeschle, and J. P. Scheidecker, Phys. Rev. A **19**, 2454 (1979).
[12] C. J. Reidl, Jr. and B. N. Miller, Phys. Rev. A **46**, 837 (1992).
[13] C. J. Reidl, Jr. and B. N. Miller, Phys. Rev. E **48**, 4250 (1993).
[14] C. J. Reidl, Jr. and B. N. Miller, Phys. Rev. E **51**, 884 (1995).
[15] T. Konishi and K. Kaneko, J.Phys.A: Math.Gen. **25**, 6283 (1992).
[16] E. Fermi, J. R. Pasta, and S. Ulam, *Collected Works of Enrico Fermi* (University of Chicago Press, Chicago, 1965), Vol. 2, p. 978.
[17] H. Hirooka and N. Saito, J. Phys. Soc. Jpn. **26** , 624 (1969).
[18] N. Saito, N. Ooyama, Y. Aizawa, and H. Hirooka, Prog. Theor. Suppl. No. 45 , 209 (1970).
[19] T. Tsuchiya, T. Konishi, and N. Gouda, Phys. Rev. E **50**, 2607 (1994).
[20] T. Tsuchiya, N. Gouda, and T. Konishi, Phys. Rev. E **53**, 2210 (1996).
[21] G. B. Rybicki, Astrophys. Space Sci. **14**, 56 (1971).
[22] A. M. Fridman and V. Polyachenko, *Physics of Gravitating Systems* (Springer-Verlag, New York, 1984).
[23] K. Ikeda, K. Matsumoto, and K. Ohtsuka, *Maxwell Bloch turbulence*, Prog. Theor. Phys. Suppl. **99** (1989), 295.
I. Tsuda, Neurocomputers and attention (A. V. Holden and V. I. Kryukov, eds.), Manchester Univ. Press, 1990.
I. Ohmine and M. Sasai, *Relaxations, fluctuations, phase transitions and chemical reactions in liquid water*, Prog. Theor. Phys. Suppl. **103** (1991), 61 – 91.
[24] H. E. Kandrup, in proceedings of the "Seventh Marcel Grossmann meeting on general relativity", ed. Ruffini, (1994), in press.

## FIGURES

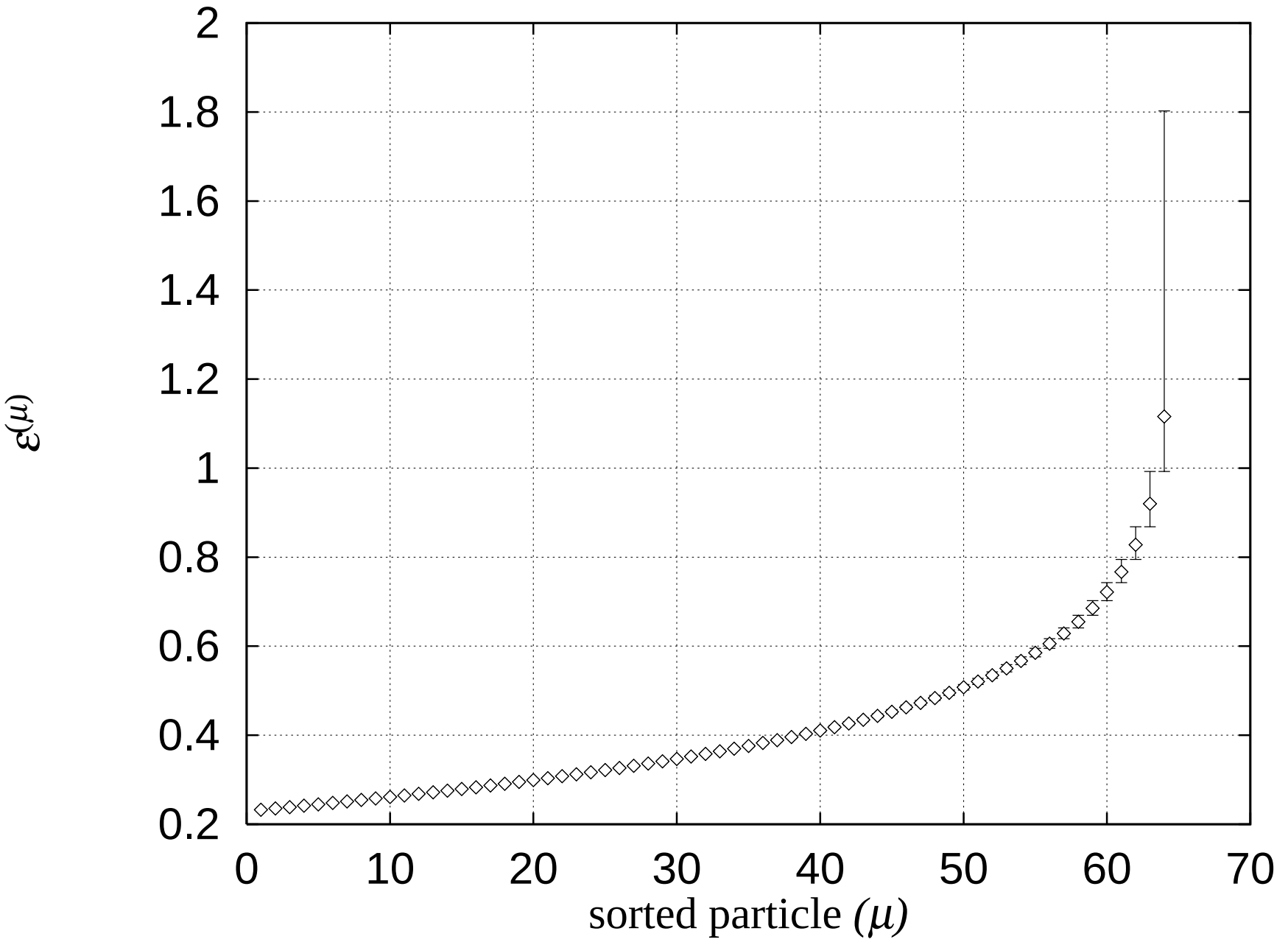

FIG. 1. Analytic solution of expectation values ($\diamond$) and variances of the energies of individual particles, which are indicated by the error bars, for the isothermal distribution.

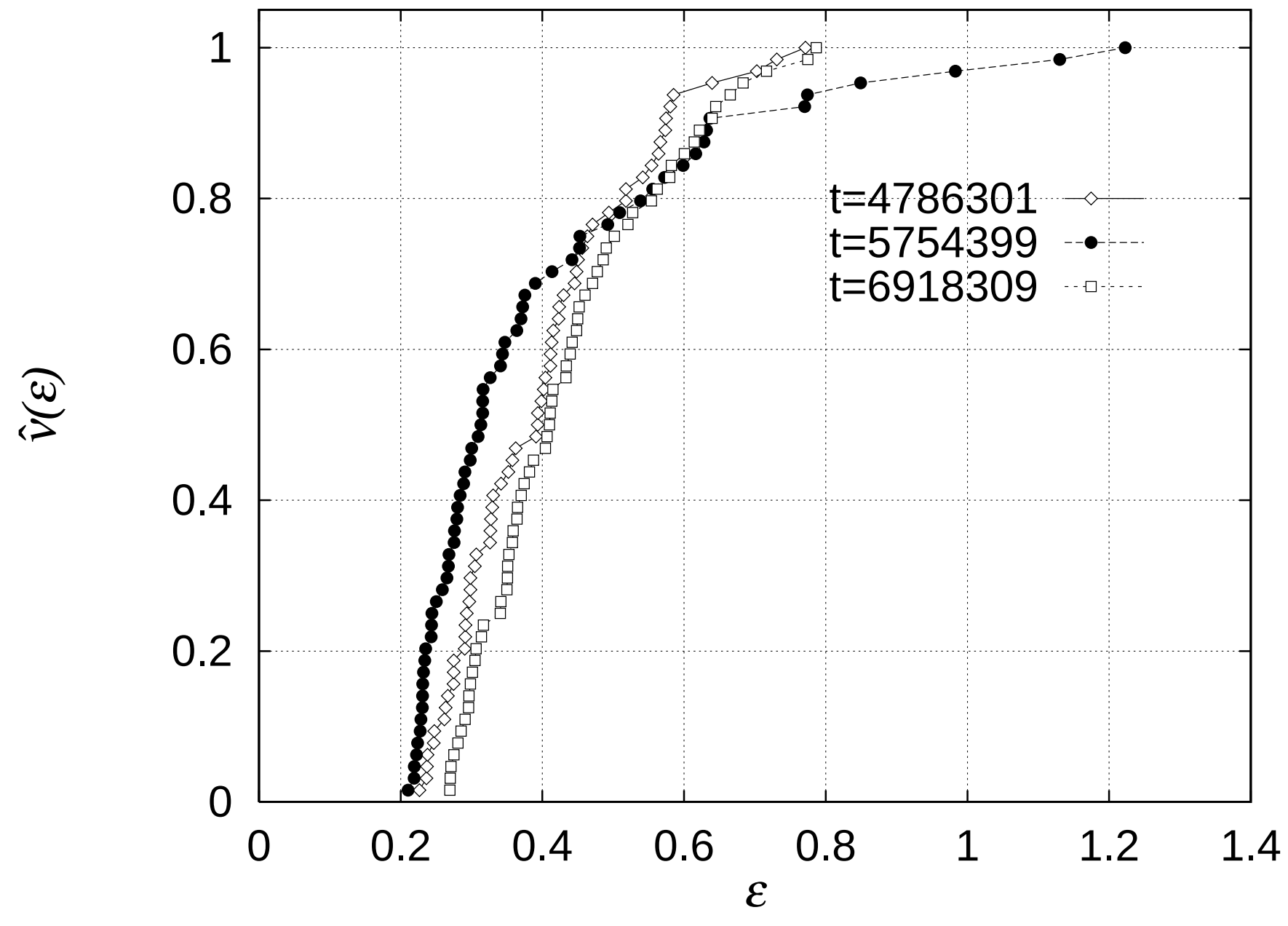

FIG. 2. Evolution of the cumulative energy distribution for a water-bag initial distribution. Three lines corresponds the distribution at the different times, which are indicated in the figure.

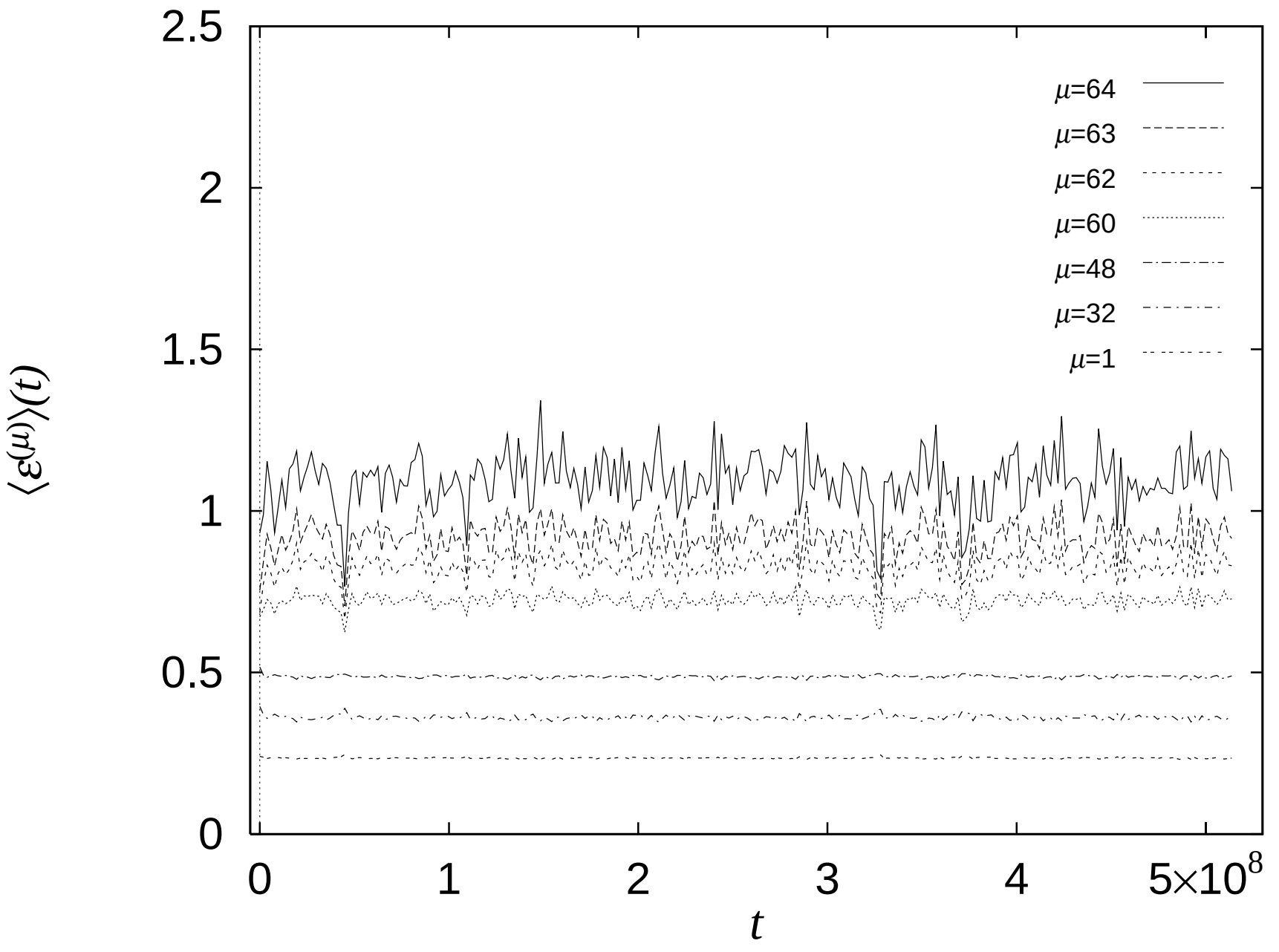

FIG. 3. Variations of the energies of the sorted particles in a long time scale. $\mu$ indicate the index of the sorted particles, such that $\mu = 1$ is the lowest energy particle and $\mu = 64$ the highest. Each variation of energy is averaged in time over the interval $\Delta t \approx 2 \times 10^5$.

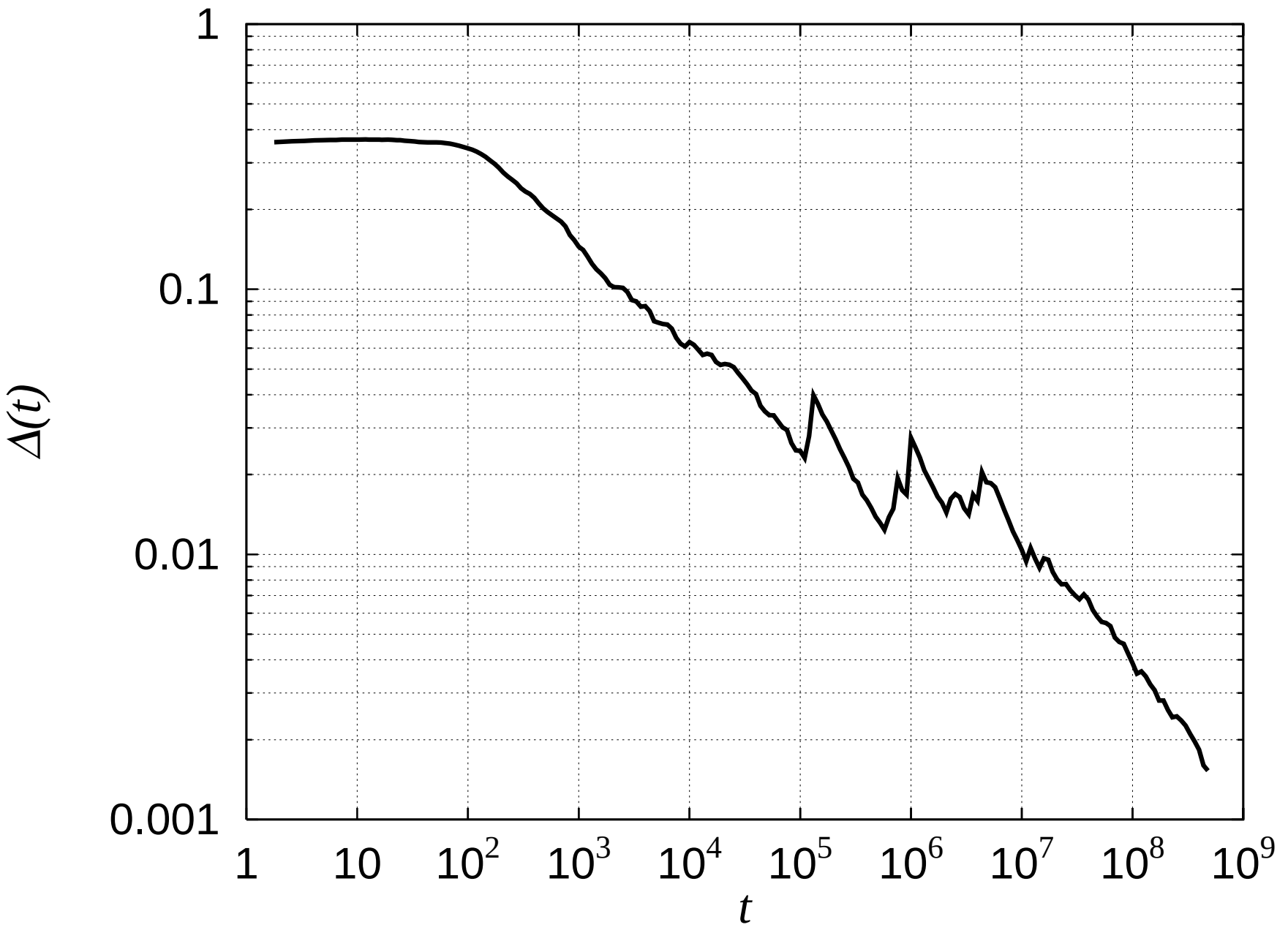

FIG. 4. Time variation of the deviations from the equipartition, $\Delta(t)$.

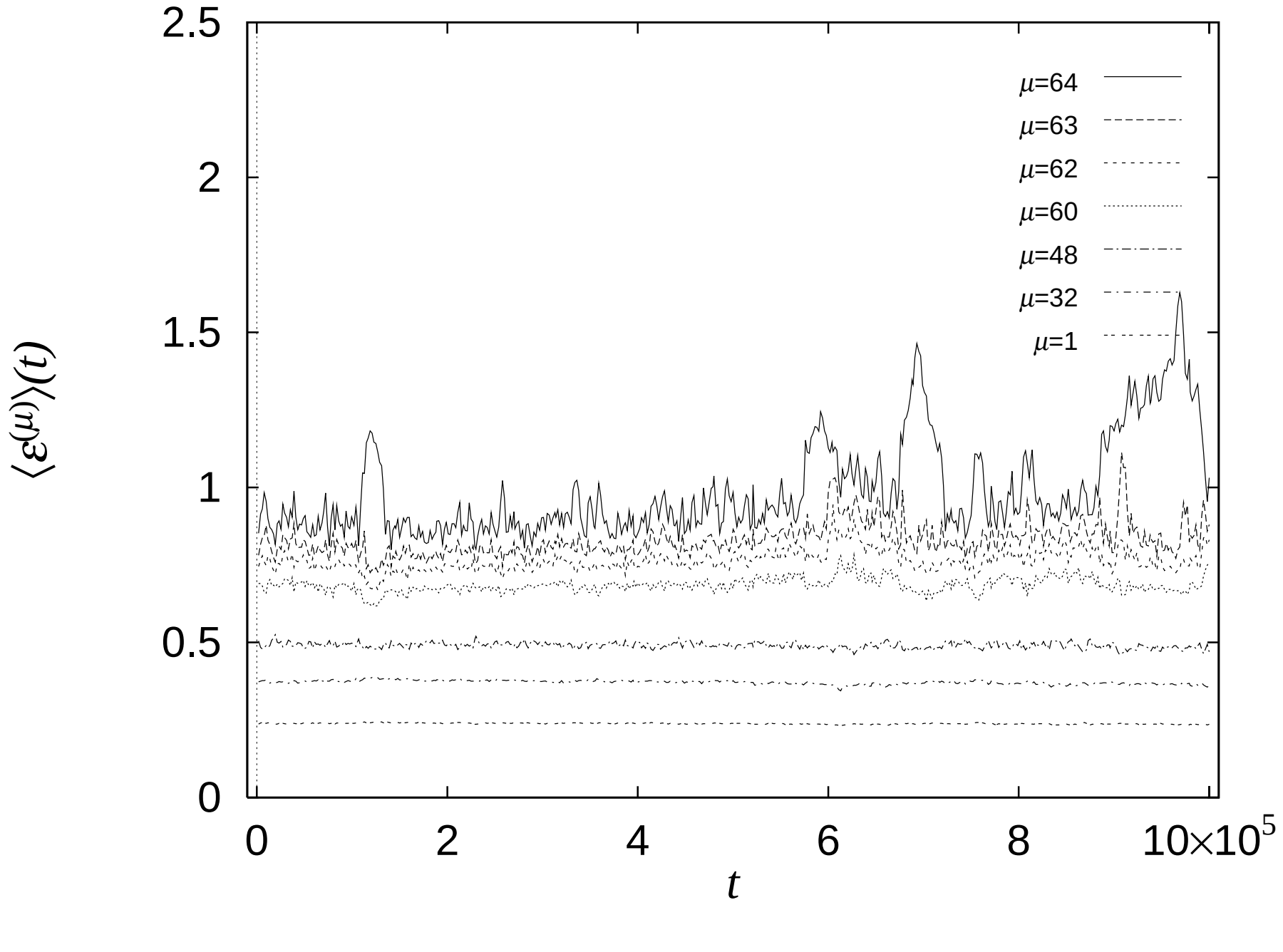

FIG. 5. Similar to Fig. 3, but only the first $10^6\, t_c$ of time in Fig. 3 is displayed. The averaging interval $\Delta t \approx 2 \times 10^3$.

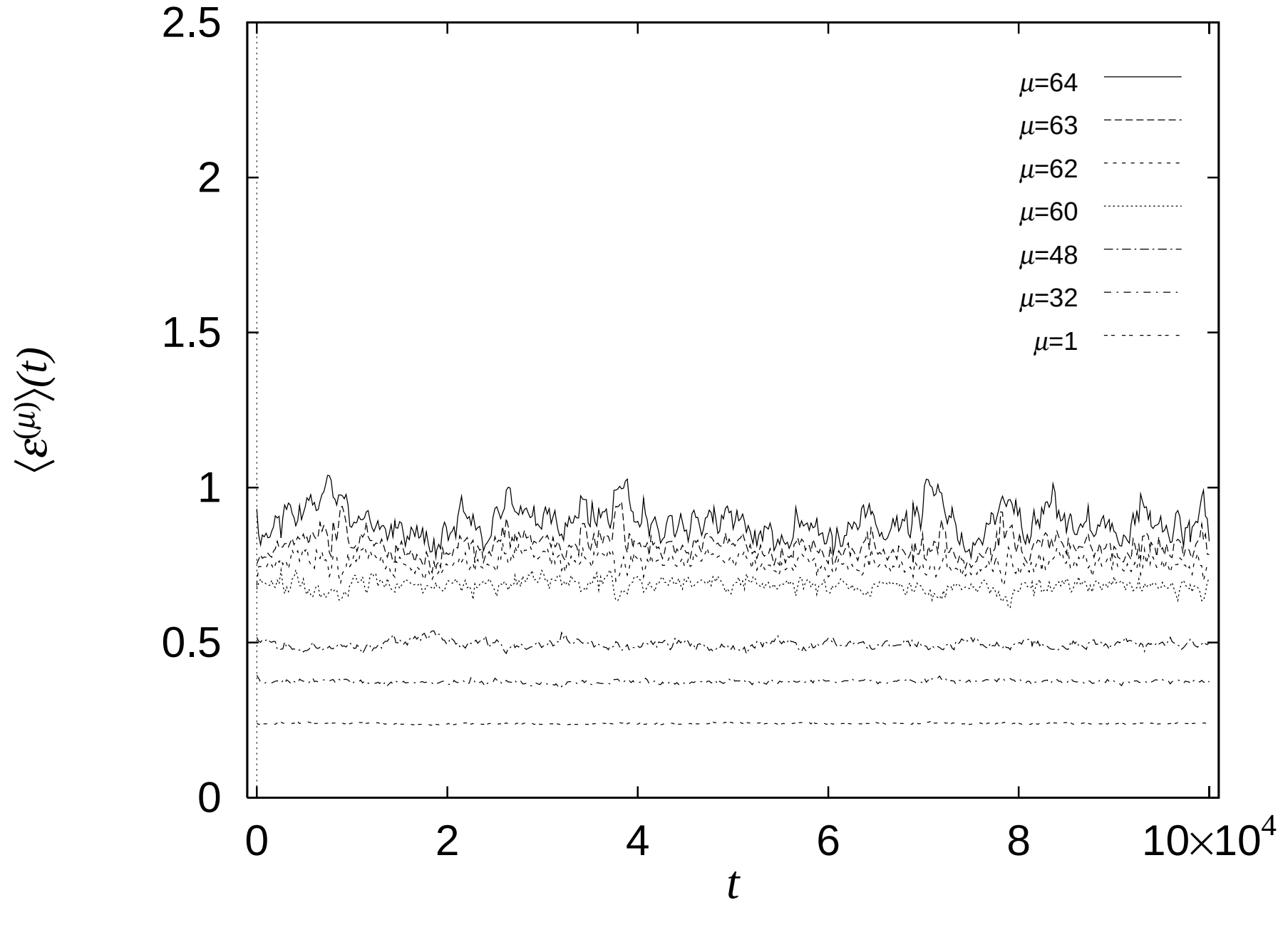

FIG. 6. Similar to Fig. 3, but only the first $10^5\, t_c$ of time in Fig. 3 is displayed. The averaging interval $\Delta t \approx 200$.

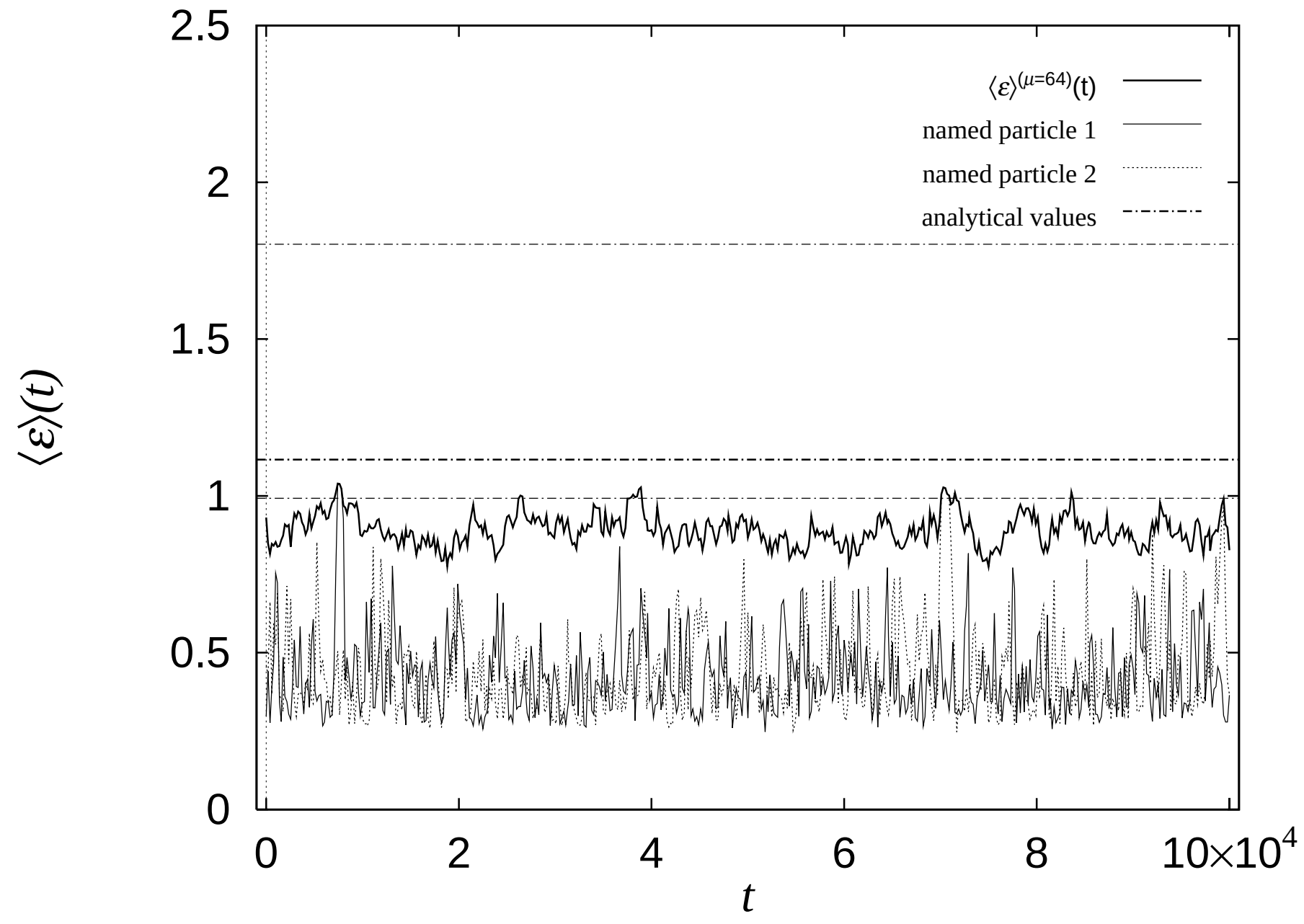

FIG. 7. Variations of the energies of the highest energy particle, $\mu = 64$, and two named particles. The energies are averaged in time over the interval $\Delta t \approx 200$. The analytic value of the expectation value and the variance of the maximum energy are shown by the bold and the faint dashed-dotted lines, respectively.

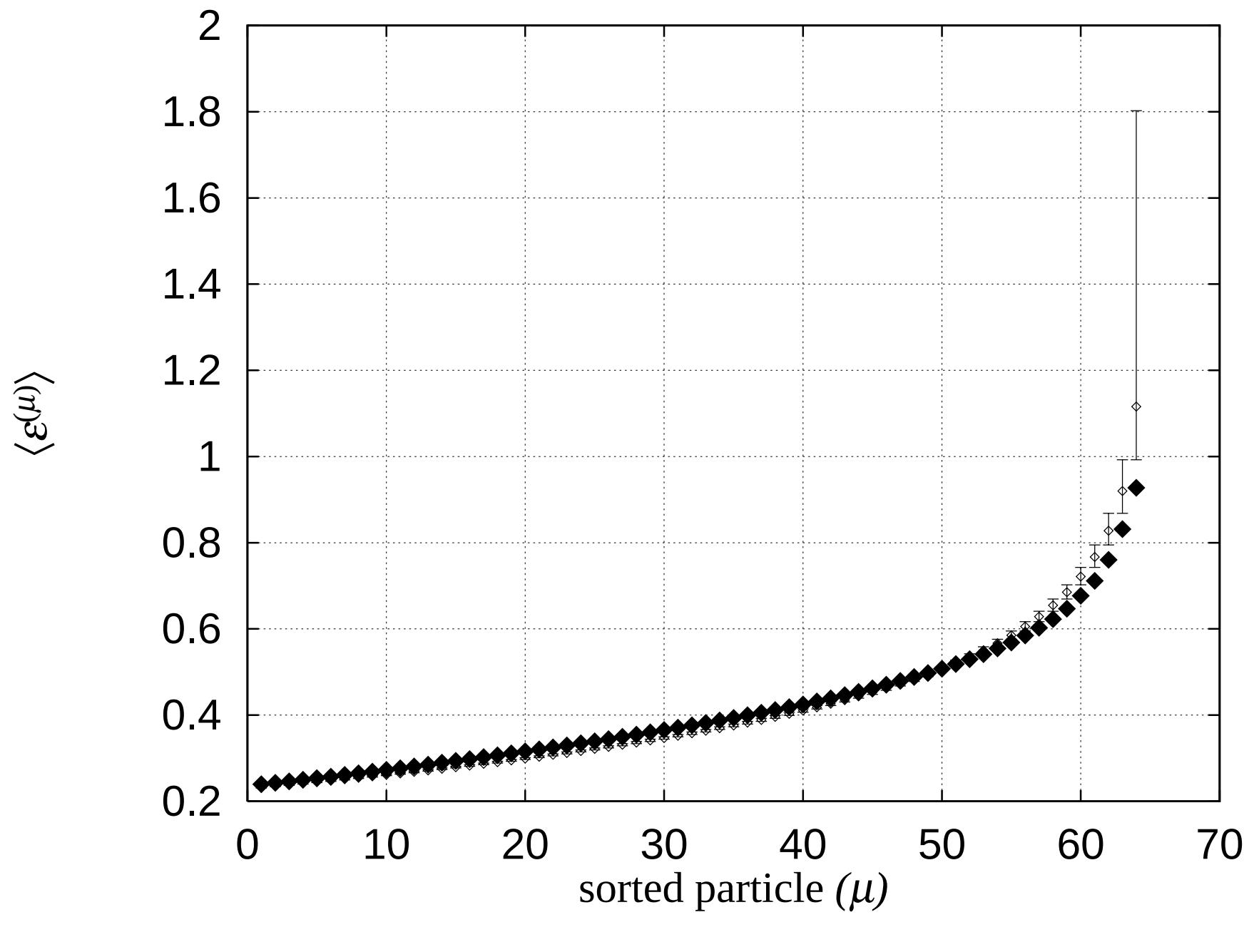

FIG. 8. The energy distribution of the sorted particles averaged over the first $10^5 t_c$. The result from the simulation is indicated by solid diamond. Analytic values and the variances of the isothermal distribution are also shown by the open diamond and the error bars, for the sake of comparison.

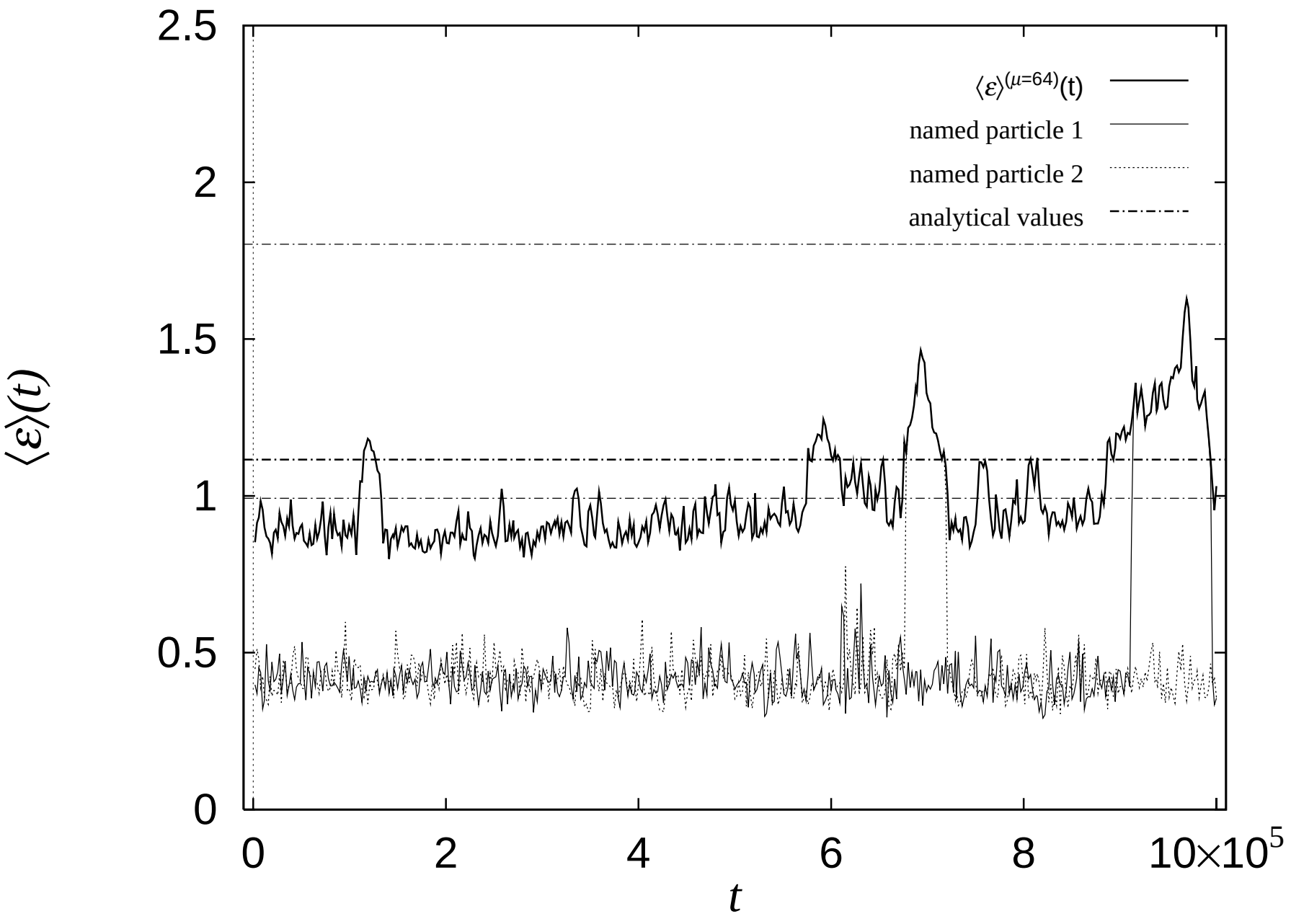

FIG. 9. Variations of the energies of the highest energy particle, $\mu = 64$, and two named particles. The energies are averaged in time over the interval $\Delta t \approx 2 \times 10^3$. The analytic value of the expectation value and the variance of the maximum energy are shown by the bold and the faint dashed-dotted lines, respectively.

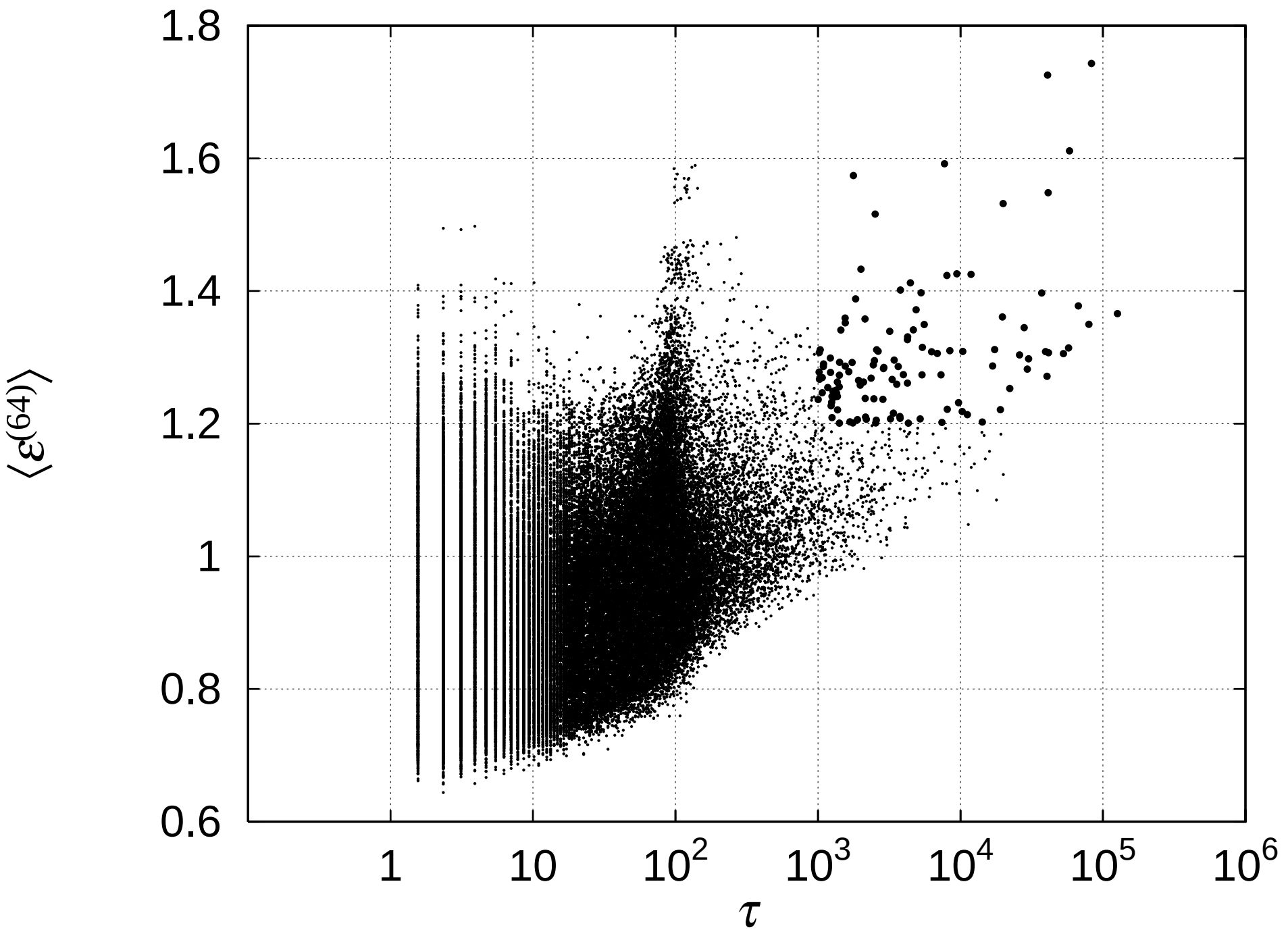

FIG. 10. Statistical distribution of the lifetime of the highest energy particle and its averaged energy. The points which lie in the region $\tau > 10^3$ and $\langle \varepsilon \rangle > 1.2$ are marked by large symbols for the sake of convenience.

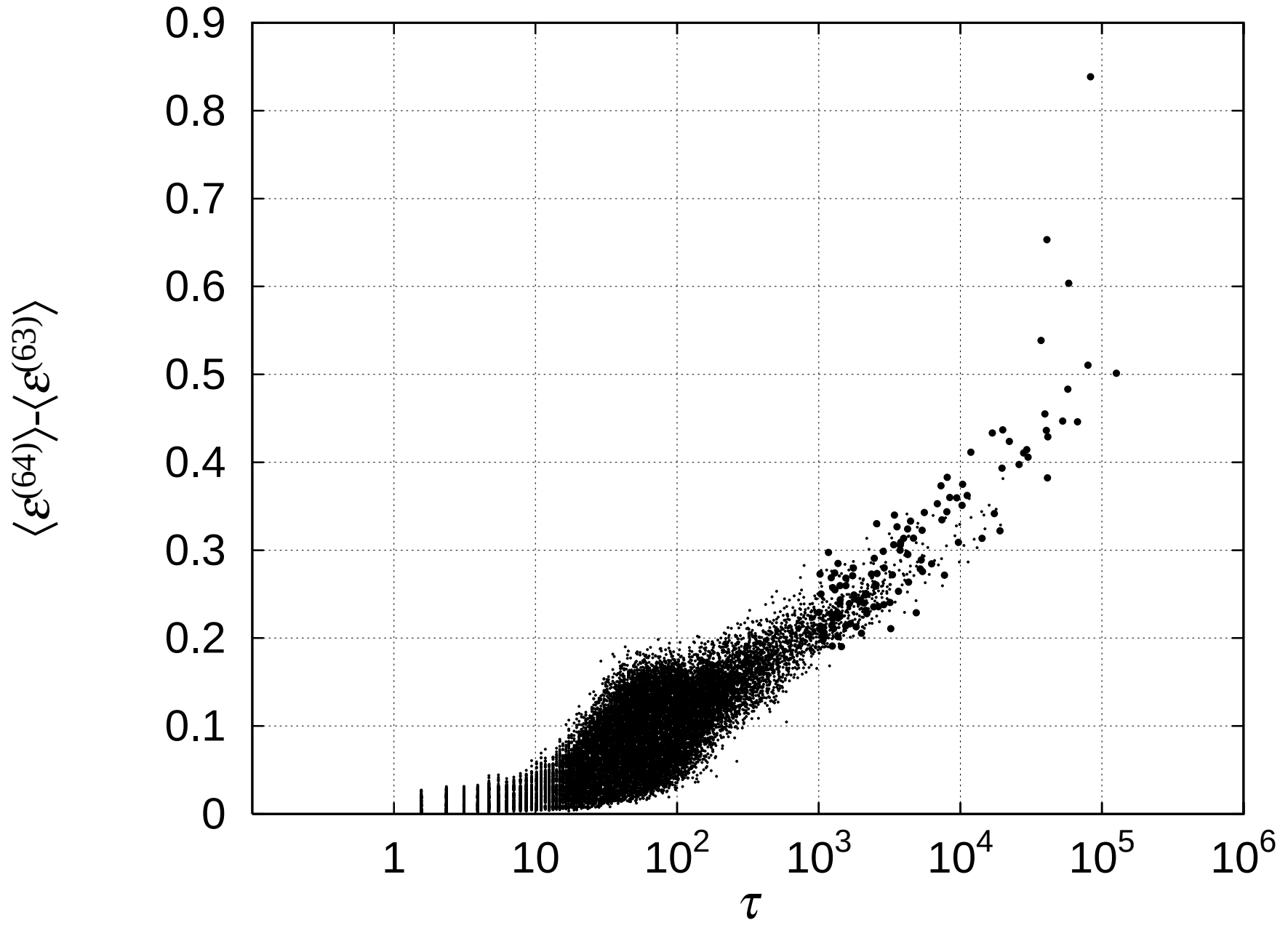

FIG. 11. Statistics distribution of the lifetime of the highest energy particle and difference of energies between the highest and the second highest.

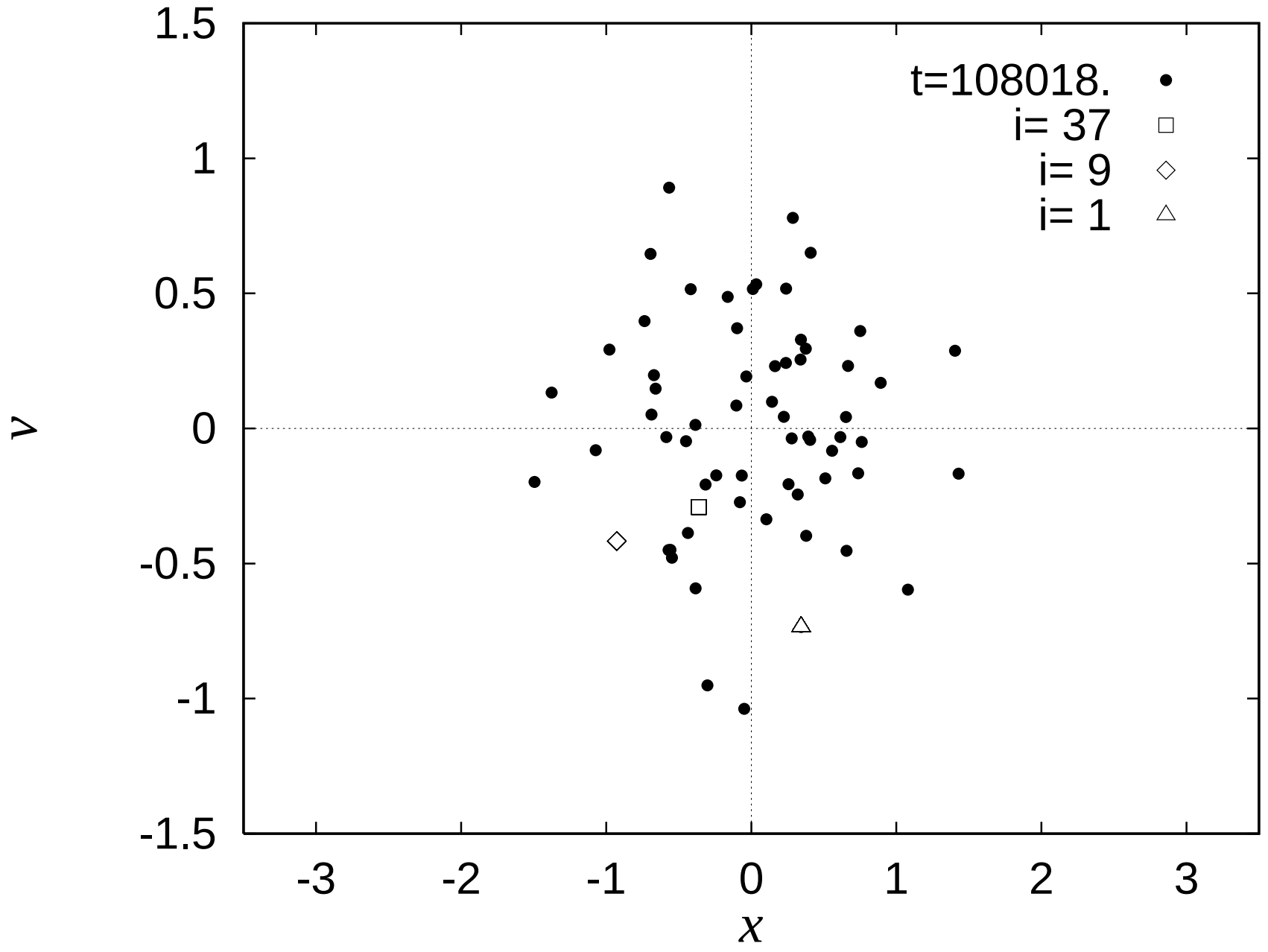

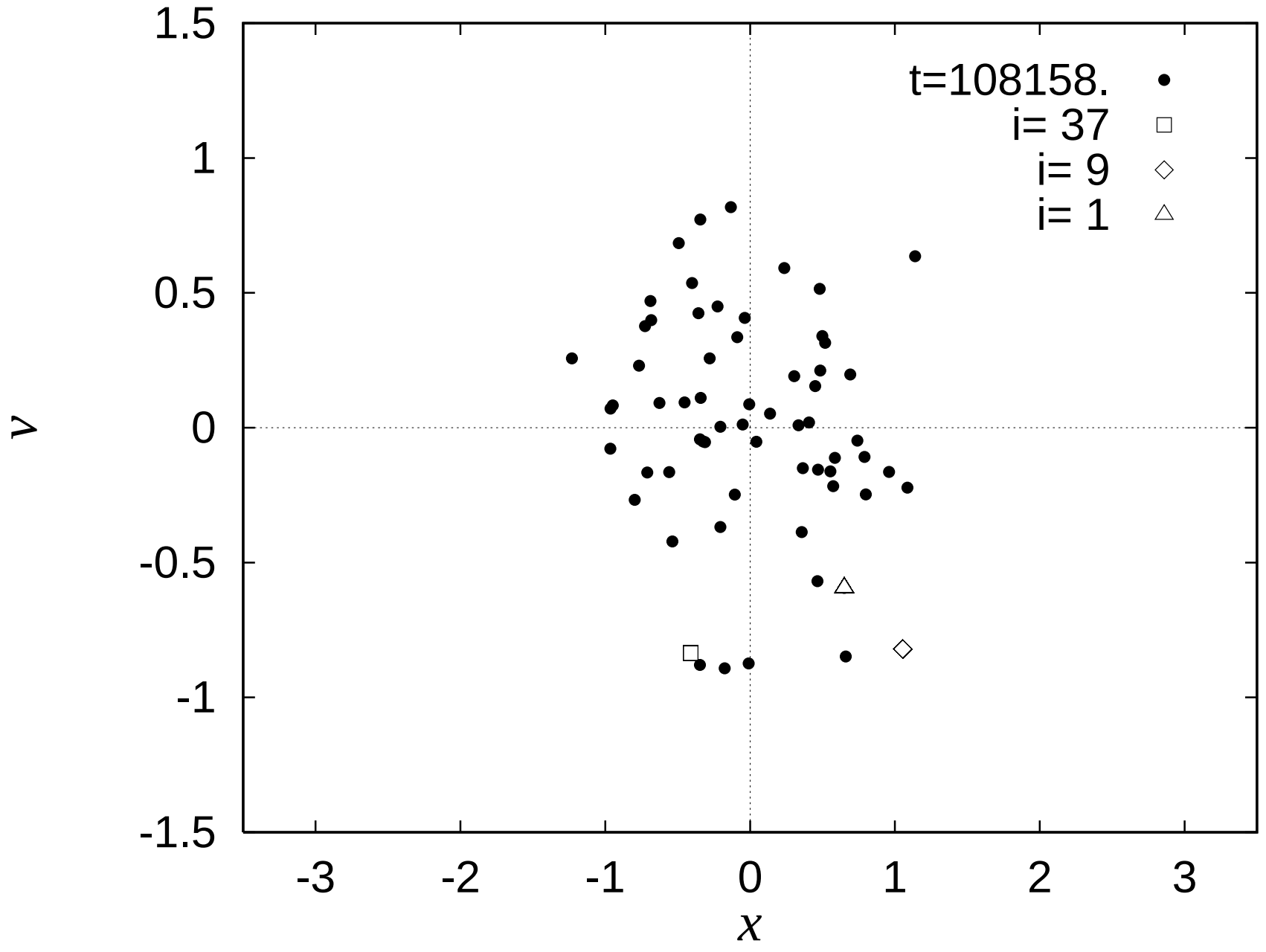

FIG. 12. A snapshot of the particle distribution of a quasiequilibrium in $\mu$-space

FIG. 13. A snapshot of the particle distribution at a beginning of a transient state in $\mu$-space

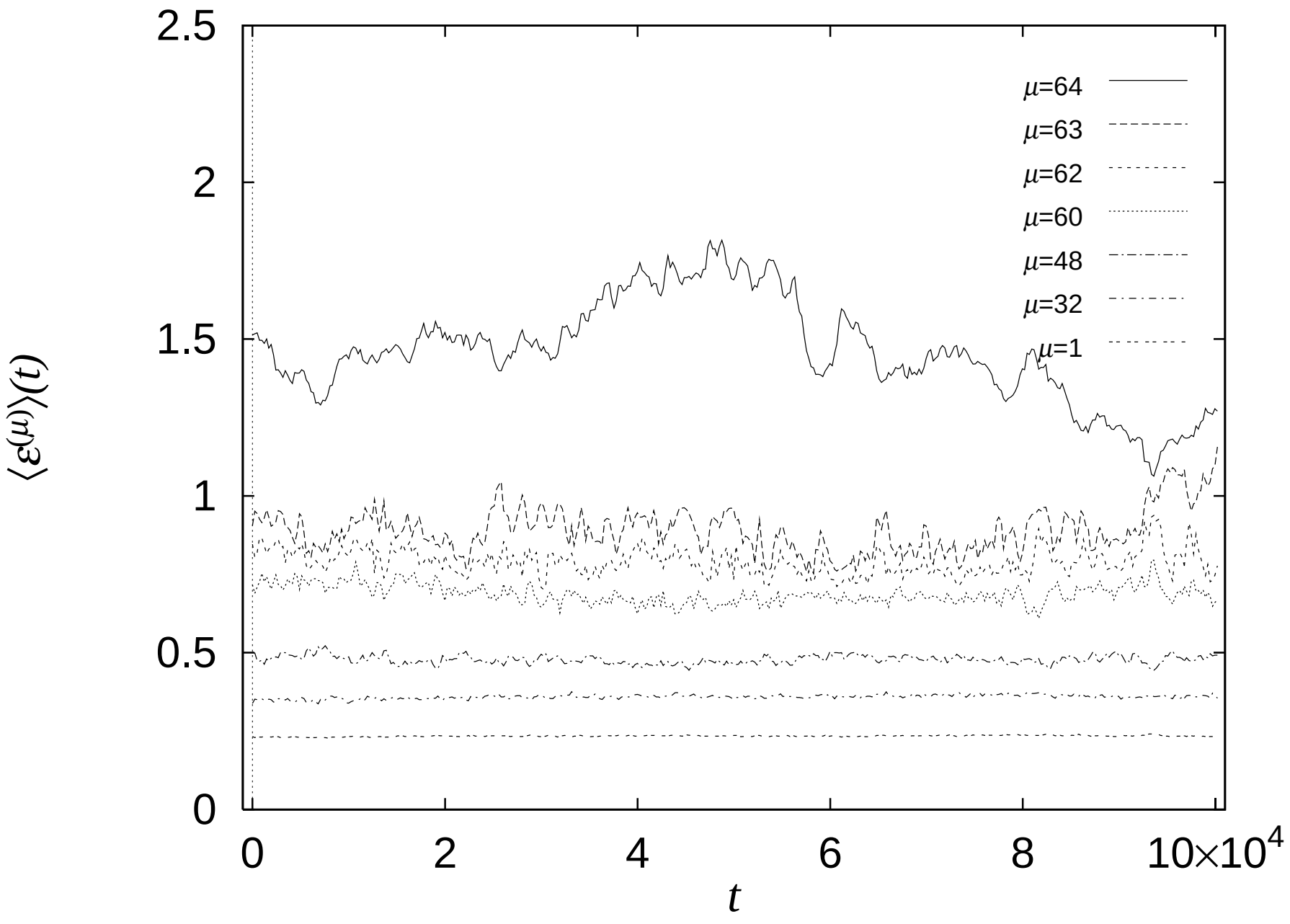

FIG. 14. Variations of the energies of the sorted particles for a transient state, in a time scale $t \sim 10^5$, and the averaging interval $\Delta t \approx 200$.

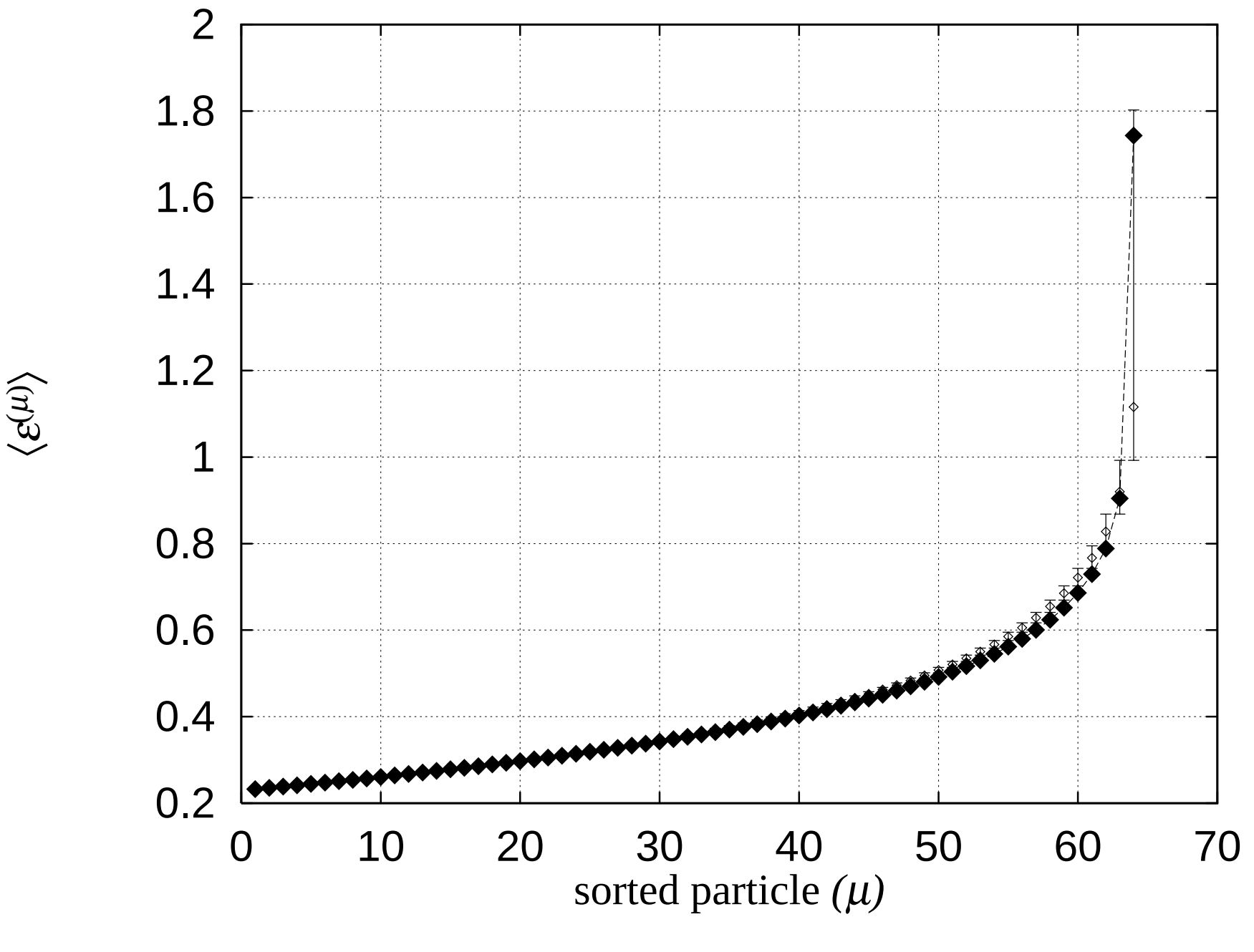

FIG. 15. Distribution of the energies in a transient state averaged over its lifetime.

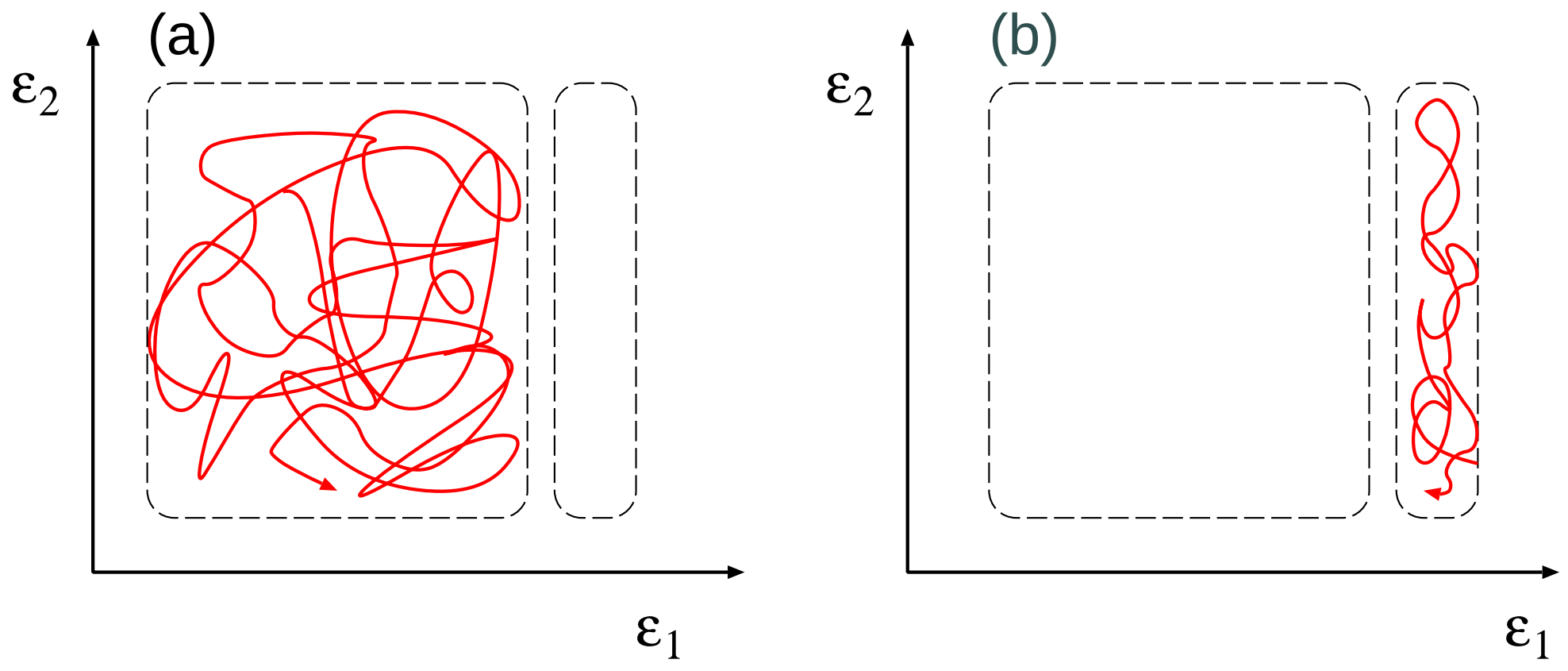

FIG. 16. Interpretation of our results: the structure of orbits in the phase space, (a) in a quasiequilibrium, (b) in a transient state. The solid line indicates an orbit, which describes the state of the system.

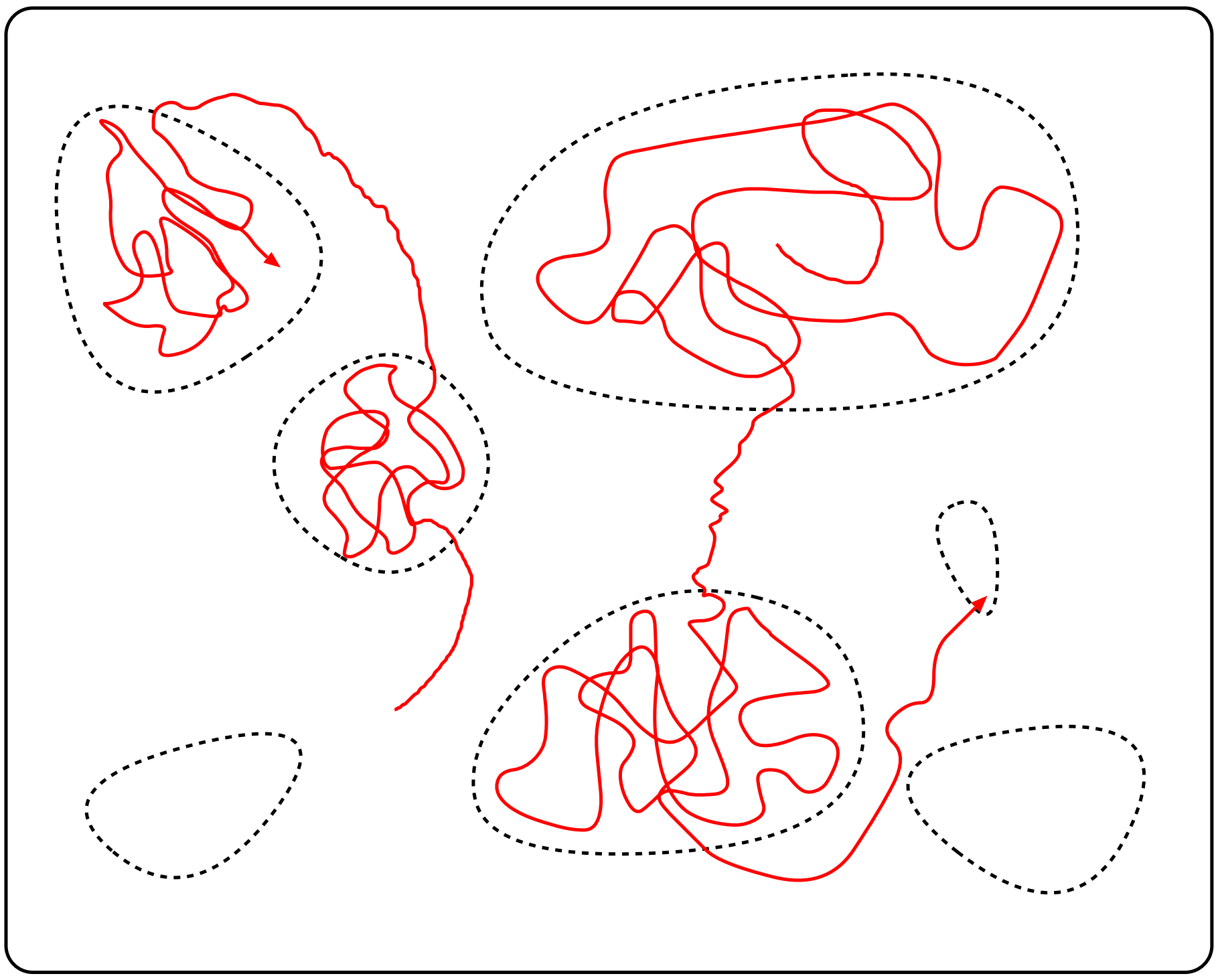

FIG. 17. Basic idea of the structure of orbits in the phase space. The solid lines indicates the orbits, which describes the states of the system. The regions which enclosed by dashed curves corresponds to the quasiequilibria. The paths which connect the quasiequilibrium regions corresponds to the transient states. Actually these structures are fractal.

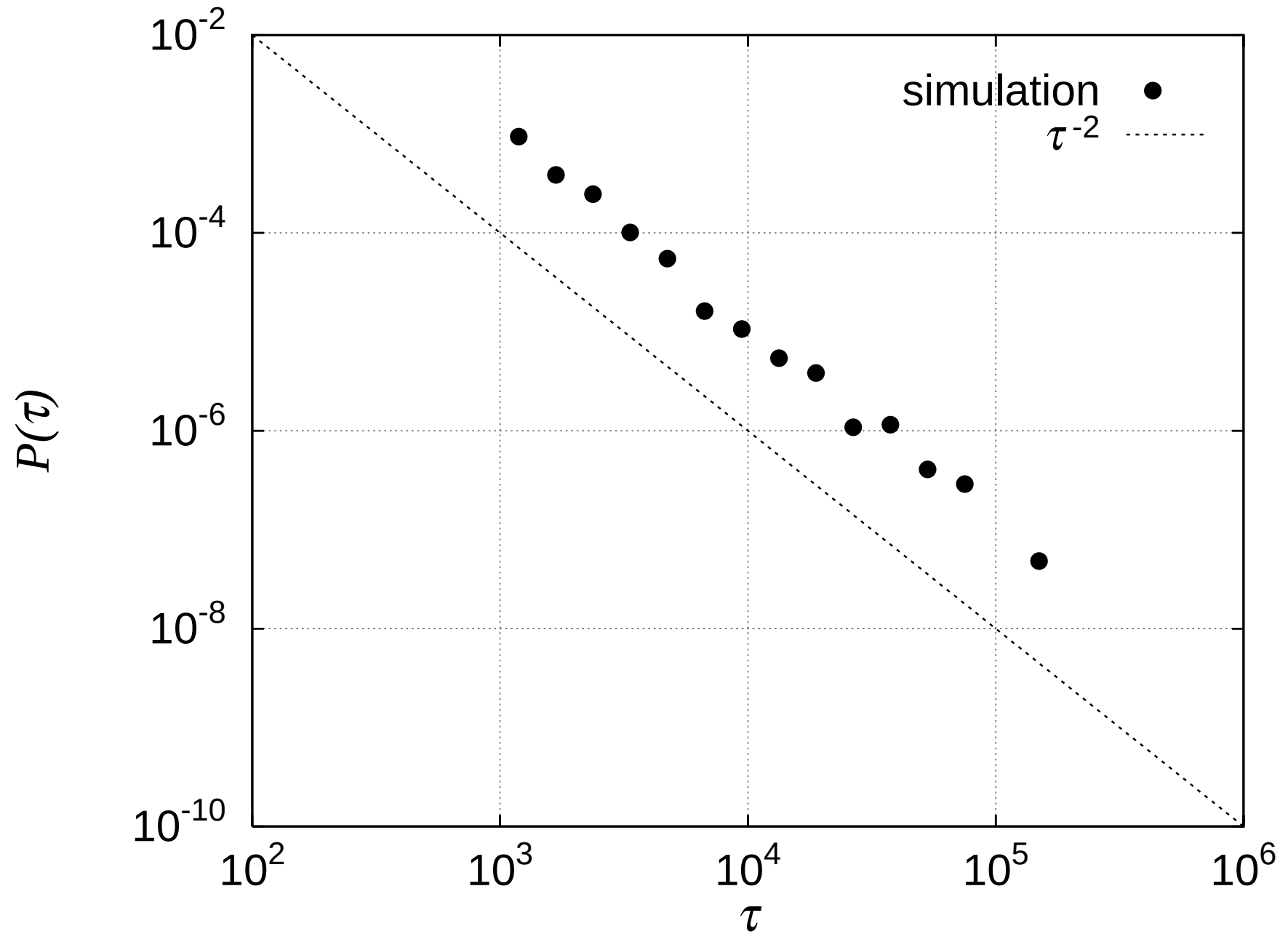

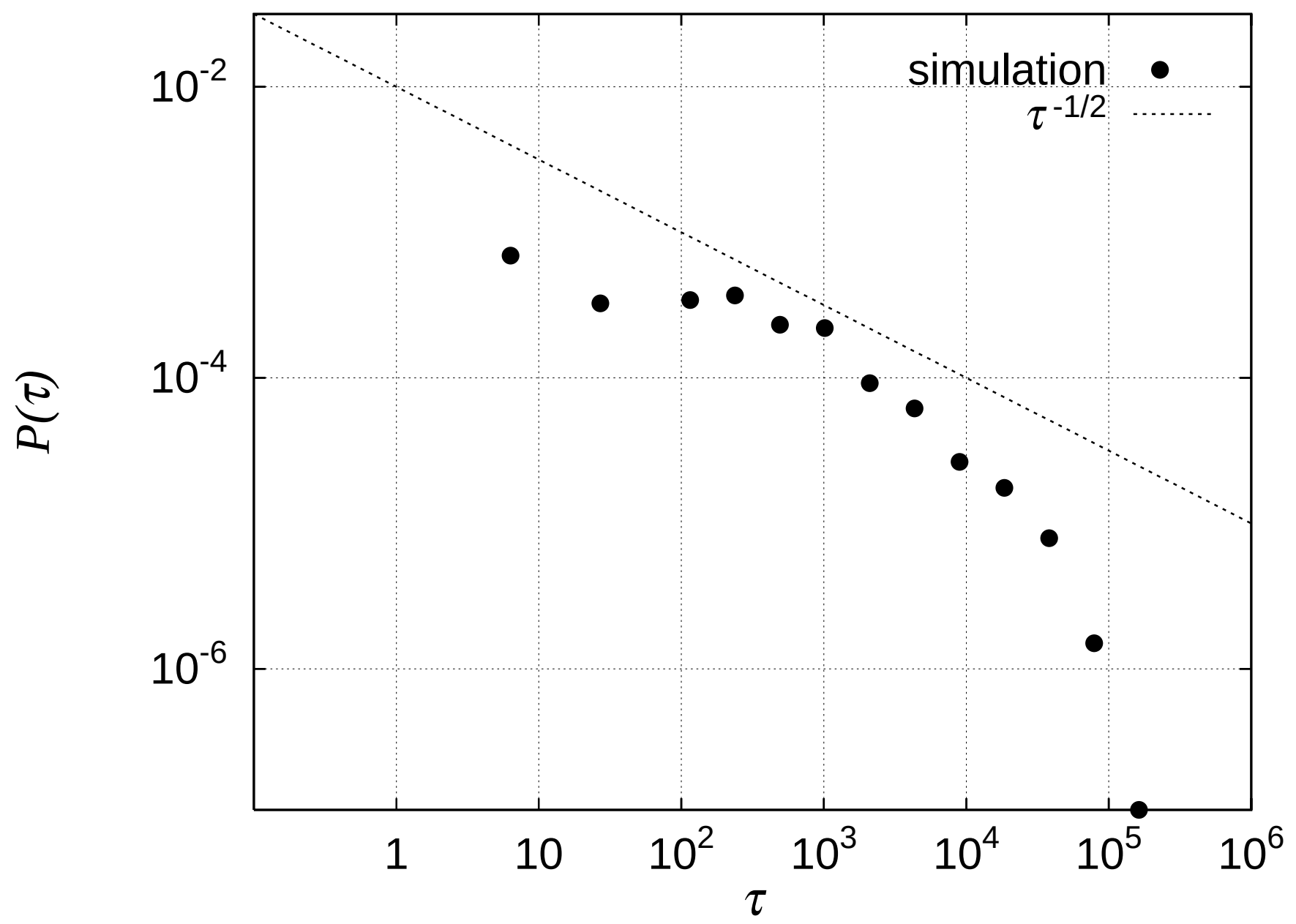

FIG. 18. Probability distribution of the lifetime of the transient state. The filled circles denotes numerical data of $T = 10^7\, t_c$, and the curve of square inverse of $\tau$ are also displayed by the solid line.

FIG. 19. Probability distribution of the lifetime of the quasiequilibria. The filled circles denotes numerical data of $T = 10^7\, t_c$, and the curve of $\tau^{-1/2}$ are also displayed by the solid line.